\documentclass[aip,reprint]{revtex4-1}
\usepackage[english]{babel}
\usepackage[utf8]{inputenc}
\usepackage[T1]{fontenc}
\usepackage[colorlinks=true,linkcolor=blue,citecolor=blue,urlcolor=blue]{hyperref}
\usepackage{bm,bbm,amssymb,amsmath}
\usepackage{mathptmx}
\usepackage{graphicx}
\usepackage{multirow,booktabs,dcolumn}
\usepackage{slashed}
\usepackage{isotope}
\usepackage{physics}
\usepackage[capitalize]{cleveref}
\usepackage{nicefrac}
\usepackage{ulem}

\usepackage[dvipsnames]{xcolor}

\begin{document}

\preprint{LA-UR-21-30239}

\title[First measurement of the $^{10}\text{B}(\alpha,n)^{13}\text{N}$ reaction in an ICF implosion at the NIF]{First measurement of the $^{10}{\rm B}(\alpha,n)^{13}{\rm N}$ reaction in an inertial confinement fusion implosion at the National Ignition Facility: Initial steps toward the development of a radiochemistry mix diagnostic}

\author{D. Lonardoni}
\email{lonardoni@lanl.gov}
\affiliation{Los Alamos National Laboratory, Los Alamos, New Mexico 87545, USA}

\author{J. P. Sauppe}
\affiliation{Los Alamos National Laboratory, Los Alamos, New Mexico 87545, USA}

\author{S. H. Batha}
\affiliation{Los Alamos National Laboratory, Los Alamos, New Mexico 87545, USA}

\author{N. Birge}
\affiliation{Los Alamos National Laboratory, Los Alamos, New Mexico 87545, USA}

\author{T. Bredeweg}
\affiliation{Los Alamos National Laboratory, Los Alamos, New Mexico 87545, USA}

\author{M. Freeman}
\affiliation{Los Alamos National Laboratory, Los Alamos, New Mexico 87545, USA}

\author{V. Geppert-Kleinrath}
\affiliation{Los Alamos National Laboratory, Los Alamos, New Mexico 87545, USA}

\author{M. E. Gooden} 
\affiliation{Los Alamos National Laboratory, Los Alamos, New Mexico 87545, USA}

\author{A. C. Hayes} 
\affiliation{Los Alamos National Laboratory, Los Alamos, New Mexico 87545, USA}

\author{H. Huang}
\affiliation{General Atomics, San Diego, California 92121, USA}

\author{G. Jungman} 
\affiliation{Los Alamos National Laboratory, Los Alamos, New Mexico 87545, USA}

\author{B. D. Keenan}
\affiliation{Los Alamos National Laboratory, Los Alamos, New Mexico 87545, USA}

\author{L. Kot}
\affiliation{Los Alamos National Laboratory, Los Alamos, New Mexico 87545, USA}

\author{K. D. Meaney}
\affiliation{Los Alamos National Laboratory, Los Alamos, New Mexico 87545, USA}

\author{T. Murphy}
\affiliation{Los Alamos National Laboratory, Los Alamos, New Mexico 87545, USA}

\author{C. Velsko}
\affiliation{Lawrence Livermore National Laboratory, Livermore, California 94550, USA}

\author{C. B. Yeamans}
\affiliation{Lawrence Livermore National Laboratory, Livermore, California 94550, USA}

\author{H. D. Whitley}
\affiliation{Lawrence Livermore National Laboratory, Livermore, California 94550, USA}

\author{C. Wilde}
\affiliation{Lawrence Livermore National Laboratory, Livermore, California 94550, USA}

\author{J. B. Wilhelmy}
\affiliation{Los Alamos National Laboratory, Los Alamos, New Mexico 87545, USA}

\begin{abstract}
We report the first measurement of the $\isotope[10]{B}(\alpha,n)\isotope[13]{N}$ reaction in a polar-direct-drive exploding pusher (PDXP) at the National Ignition Facility (NIF). 
This work is motivated by the need to develop alternative mix diagnostics, radiochemistry being the focus here.
The target is composed of a $65/35\,\rm at.\,\%$ deuterium-tritium (DT) fill surrounded by a roughly $30\,\mu\rm m$ thick beryllium ablator. 
The inner portion of the beryllium ablator is doped with $10\,\rm at.\,\%$ of \isotope[10]{B}. 
Radiation-hydrodynamics calculations were performed in 1D to optimize both the remaining boron rho-R and the DT neutron yield. 
A charged-particle transport post-processor has been developed to study $\alpha$-induced reactions on the ablator material. 
Results indicate a large \isotope[13]{N} production from $\alpha$-induced reactions on \isotope[10]{B}, measurable by the radiochemical analysis of gaseous samples system at the NIF. 
The PDXP target N201115-001 was successfully fielded on the NIF, and nitrogen from the $\isotope[10]{B}(\alpha,n)\isotope[13]{N}$ reaction was measured. 
The \isotope[13]{N} production yield, as well as the DT neutron yield, was, however, lower than expected.
Some of the reduced yields can be explained by the oblate shape, but the ratios of the various radiochemical signals are not commensurate with expectations based on a simple reduction of the 1D results.
Preliminary 2D radiation-hydrodynamics computations are consistent with the experimental measurements, and work is ongoing to extend the radiochemistry analysis into higher dimensions.
\end{abstract}

\maketitle

\section{Introduction}
\label{sec:intro}
Inertial confinement fusion (ICF)\cite{Nuckolls:1972} involves a target capsule filled (usually) with deuterium-tritium (DT) fuel that is compressed and heated by the energy delivered to the outer ablator layer of the target using either direct laser light,\cite{Craxton:2015} or x-rays generated inside a \textit{Hohlraum} illumined by lasers.\cite{Lindl:1995} Much of the outer layer ablates off, causing the remaining target to implode via the rocket effect. The central fuel is compressed and heated as the remaining ablator material moves inwards, and the ablator's kinetic energy is converted into fuel internal energy. As pressure builds in the DT gas, the ablator begins decelerating and eventually stagnates, representing the point at which $PdV$ work ceases to increase the fuel energy.

During the deceleration phase of the implosion, Rayleigh-Taylor\cite{Rayleigh:1882,Taylor:1950} (RT) instabilities can form, leading to mixing between the ablator material and the DT fuel.\cite{Palaniyappan:2020}
The shock waves involved in the implosion can induce Richtmyer-Meshkov\cite{Richtmyer:1960,Meshkov:1969} (RM) instabilities, possibly contributing to mixing as well.\cite{Bose:2015} 
RT, RM, and other instabilities are seeded by surface finish imperfections, illumination nonuniformity, the fill tube, glue spot, and the capsule support tent.\cite{Clark:2016,Haines:2017,Weber:2017,Clark:2019,Haines:2019,Weber:2020} 
Mixing between the ablator material and the DT gas is undesirable, because it reduces the fuel temperature. It also raises the adiabat of the fuel,\cite{Cheng:2016,Cheng:2018} making it harder to compress. 
Both of these effects reduce the performance of the capsule. 

One method for diagnosing the mix of ablator material into the DT fuel is to study the degree of elevated x-ray emission from the burning DT fuel.\cite{Ma:2013}
X-ray imaging can also show distortions during the implosion and the hot spot at peak compression.\cite{Doppner:2020} 
However, there are capsule designs being developed for the National Ignition Facility\cite{Edwards:2013} (NIF), such as double shell and pushered single shell,\cite{Montgomery:2018,Dewald:2019} for which this x-ray technique cannot be used effectively because the designs involve a high-$Z$ shell surrounding the DT fuel that is opaque to x-rays. 

A possible alternative mix diagnostic involves measuring the interaction of alpha particles, produced in the ${\rm D}+{\rm T}\rightarrow \alpha\,(3.5\,{\rm MeV}) +n\,(14.1\,{\rm MeV})$ reaction, with the ablator material.\cite{Colvin2008} 
The Radiochemical Analysis of Gaseous Samples (RAGS) facility \cite{Shaughnessy:2012} at NIF, which collects the gaseous debris emitted into the chamber from imploded capsules, allows for quantitative measurements of $\alpha$-induced reactions, as long as the reaction products are gaseous. 
If mixing occurs, the DT $\alpha$ production is reduced due to the decrease of the capsule performance. 
The reduction in the DT yield from mixing may be partially compensated for by the closer proximity of the ablator materials to the alpha production region. 
The net effect is expected to be an appreciable change in the radiochemical signature ratios, making radiochemistry (RadChem) a valuable mix diagnostic for ICF studies.\cite{Colvin2008}

\begin{figure}[t]
\includegraphics[width=\linewidth,trim=0.0cm 0.6cm 0.0cm 1.0cm, clip=true]{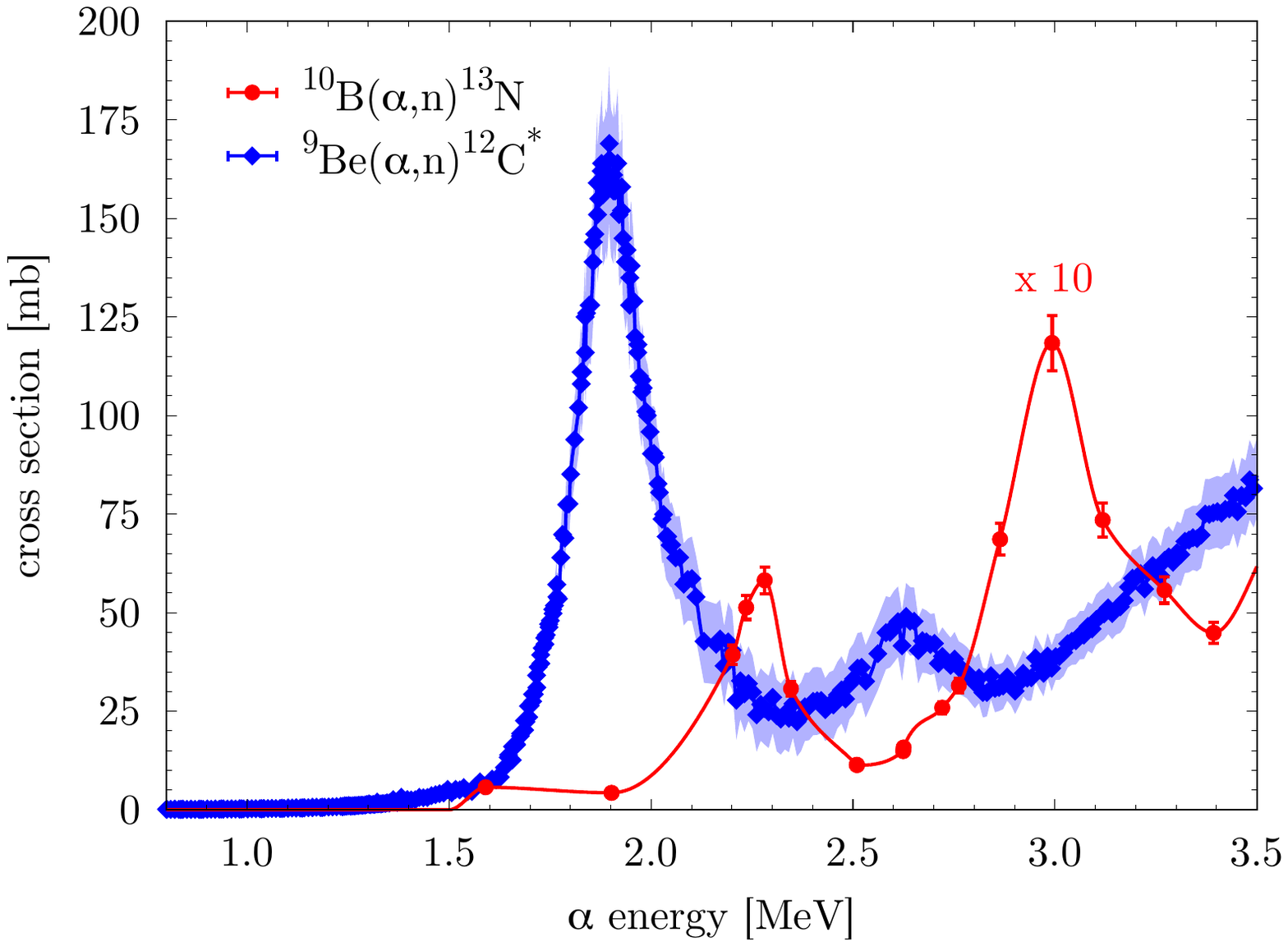}
\caption[]{Experimental cross sections as a function of the incident $\alpha$-particle energy (laboratory reference frame). 
Red dots and error bars for $\isotope[10]{B}(\alpha,n)\isotope[13]{N}$ (ground-state transition) from Ref.~\onlinecite{Liu:2019}.
Blue diamonds and error bands for $\isotope[9]{Be}(\alpha,n)\isotope[12]{C^*}(\gamma\;@\;4.4\,{\rm MeV})$ from the EXFOR database.\cite{Otuka:2014} 
Solid lines are interpolations on the experimental results.}
\label{fig:cross}
\end{figure}

In this work, we present the first measurement of the $\isotope[10]{B}(\alpha,n)\isotope[13]{N}$ reaction in an ICF implosion, using a polar-direct-drive exploding pusher (PDXP) shot at NIF, N201115-001, as a first step in the development of RadChem mix diagnostics. 
The cross section for this reaction is shown in \cref{fig:cross} (red circles), as discussed in Ref.~\onlinecite{Liu:2019}. 
There is a significant peak in the cross section for $\alpha$ particles with $\sim3\,\rm MeV$ of energy, slightly below the expected energy for $\alpha$'s produced from the D+T fusion reaction.

The N201115-001 target consisted of a 65/35\,at.\,\% deuterium-tritium fill ($8\,\rm atm$ at room temperature) surrounded by a boron-doped beryllium ablator. 
In addition to the \isotope[13]{N} measurement, RAGS was also used to measure the neutron-induced reaction on argon, $\isotope[40]{Ar}(n,\gamma)\isotope[41]{Ar}$, providing a potential additional diagnostic for the areal density of the remaining ablator material at the time of the DT burn, in a similar manner as Ref.~\onlinecite{Wilson:2017}.
As described later in the manuscript, the \isotope[41]{Ar} signal can also be used as a mix metric together with \isotope[13]{N}.
(Argon was present in the beryllium ablator as a by-product of the sputter-coating fabrication process.) 
The gamma reaction history (GRH) capability\cite{Herrmann:2010} was also employed in an attempt to measure $4.4\,\rm MeV$ gamma from the de-excitation of carbon $2^+$ excited states coming from the $\isotope[9]{Be}(\alpha,n)\isotope[12]{C^*}$ reaction, as a cross calibration for the RadChem diagnostic.
This reaction cross section is also shown in \cref{fig:cross} (blue diamonds).
The capsule produced $4.04\cdot10^{14}$ DT neutrons, and $7\cdot10^{6}$ \isotope[13]{N} and $10^{5}$ \isotope[41]{Ar}. 
A combination of low signal and competition between the $4.4\,\rm MeV$ and the DT fusion gamma rays allowed only an upper limit of $2\cdot10^{12}$ of the $4.4\,\rm MeV$ gamma ray yield (see \cref{tab:exp} and the discussion in Sec.~\ref{sec:exp}). Due to the ablator being made primarily of beryllium, the GRH signal was not able to infer a shell $\rho R$ from the data as has been done for carbon-based ablators.\cite{Meaney:2021}

\begin{table}[b]
\caption{Summary of the experimental results for the N201115-001 PDXP shot. The last column reports the source/detector used for the estimates.
RAGS and GRH signal thresholds are also shown in parenthesis.}
\label{tab:exp}
\begin{ruledtabular}
\begin{tabular}{lcc}
\isotope[10]{B} dopant level    & $10\,\rm at.\,\%$        & General Atomics \\
\isotope[40]{Ar} concentration  & $0.3\,\rm at.\,\%$       & General Atomics \\
Bang time                       & $4.5\pm0.5\,\rm ns$         & GRH\cite{Herrmann:2010} \\
                                & $4.83\pm0.13\,\rm ns$       & SPIDER\cite{Khan:2012} \\
Burn width                      & $600\pm75\,\rm ps$          & GRH \\
                                & $618.50\pm270.59\,\rm ps$   & SPIDER \\
DT$n$ yield                     & $4.040\pm0.126\cdot10^{14}$ & Well-NAD\cite{Yeamans:2016} \\
\isotope[13]{N} yield           & $7.0\pm1.4\cdot10^{6}$\;\;$(Y_\text{th}=10^{5})$   & RAGS\cite{Shaughnessy:2012} \\
\isotope[41]{Ar} yield          & $\sim1.0\cdot10^{5}$\;\;$(Y_\text{th}=10^{5})$     & RAGS \\
$\gamma$ $(4.4\,\rm MeV)$ yield & $<2\cdot10^{12}$\;\;$(Y_\text{th}=4\cdot10^{10})$  & GRH \\
\end{tabular}
\end{ruledtabular}
\end{table}

The remainder of the paper is organized as follows. In Sec.~\ref{sec:target}, we discuss the target design, the radiation-hydrodynamics modeling framework and initial optimization studies, and the target fabrication process.
In Sec.~\ref{sec:analysis}, we present tuned post-shot simulations in 1D, the RadChem post-processing framework is introduced, and we discuss sensitivities observed from applying this methodology to our 1D simulations.
The experimental results are presented in Sec.~\ref{sec:exp}, and more detailed 2D radiation-hydrodynamics modeling is also considered here. 
Finally, we conclude in Sec.~\ref{sec:concl} and discuss next steps for the platform.

\section{Initial design studies}
\label{sec:target}
We based our original design on a previously developed low-convergence, high-yield polar direct-drive exploding pusher platform.\cite{Ellison:2018} 
The particular capsule we chose to emulate was fielded on NIF shot N190707-001 (colloquially known as ``Little Guy''), a CH ablator target that yielded $4.8\times 10^{15}$ neutrons.\cite{Yeamans:2021}
In that design, the laser drive was optimized to minimize multi-dimensional effects, enabling physics studies in an approximately 1D configuration. 
Further, this laser drive (total of $585\,\rm kJ$ laser energy) has the advantage of inducing no known optics damage at NIF, providing a robust neutron signal with no risk to the laser. 
In this design work, we opted to keep the same laser pulse history and beam pointings due to the record performance achieved with no optics damage.

In order to measure the primary $\isotope[10]{B}(\alpha,n)\isotope[13]{N}$ reaction, the capsule design had to be modified to include \isotope[10]{B} near the DT fuel region, placing it in close proximity to the $\alpha$ particles produced from thermonuclear burn. 
However, it proved difficult to fabricate leak-tight pure boron ablators that can hold DT gas at the required fill pressures, limiting our design choices to only utilizing a doped \isotope[10]{B} region. 
In a CH ablator, \isotope[13]{N} can also be produced from the competing reactions $\isotope[12]{C}(d,n)\isotope[13]{N}$ and $\isotope[13]{C}(p,n)\isotope[13]{N}$. 
This background production from carbon can potentially be comparable to the production from boron,\cite{Liu:2019} posing significant technical challenges to the extraction of the $\alpha$-induced contribution from the total signal.
Instead, we considered a beryllium ablator with a region of boron-doped beryllium on the inner part of the capsule. 
This has the additional advantage of providing a complementary diagnostic via the measurement of $4.4\,\rm MeV$ gammas coming from the $\isotope[9]{Be}(\alpha,n)\isotope[12]{C^*}(\gamma\;@\;4.4\,{\rm MeV})$ reaction.

We note that the relevant reactions for this study are only those populating the \isotope[13]{N} lowest-energy state, the angle-integrated cross section of which has been recently remeasured in the energy range relevant for DT nuclear burn\cite{Liu:2019} (see \cref{fig:cross}). 
In fact, the ground-state of \isotope[13]{N} has a half-life of $t_{1/2}=9.965(4)\,\rm min$,\cite{Ajzenberg:1991} long enough to allow collection using RAGS. 
\isotope[13]{N} excited states are proton unbound and will, instead, immediately decay to stable \isotope[12]{C}.

The long half-life of \isotope[41]{Ar} ($t_{1/2}=109.61(4)\,\rm min$\cite{ENSDF}) makes it a good candidate for collection at RAGS, allowing us to also measure the reaction $\isotope[40]{Ar}(n,\gamma)\isotope[41]{Ar}$. 
The sputter-coating process used to fabricate these capsules results in a small amount of argon being uniformly deposited within the ablator as well. 
From our analysis of the fabricated capsule, the \isotope[40]{Ar} concentration in the ablator shell was estimated to be of the order of $0.3\,\rm at.\,\%$. 
Indirect-drive targets are sensitive to the argon concentration, primarily as a result of the modification in the opacity of the ablator.\cite{Wilson:2017}
However, this is not as well studied for direct-drive targets. Preliminary modeling, discussed in Sec.~\ref{sec:sensitivity}, finds sensitivity to the argon concentration.

\subsection{Modeling framework and design optimization}
\label{sec:preshot}
Our design calculations are carried out with \texttt{xRAGE},\cite{Gittings:2008,Haines:2017} the Los Alamos National Laboratory's Eulerian radiation-hydrodynamics code. 
A laser package has recently been implemented for 1D and 2D \texttt{xRAGE} computations.\cite{Marozas:2018,Haines:2020}
Here, we use an \textit{ad hoc} laser power multiplier, $\eta_{\rm laser}$, to reduce the input laser energy to account for cross-beam energy transfer (CBET) or other laser plasma instabilities (LPI) that are not currently modeled. 
(Although a CBET package is available, it increases the computational cost considerably and was not used in our preliminary design calculations.)
Thermal conduction in \texttt{xRAGE} (and many radiation-hydrodynamics codes) is handled using the Spitzer-Harm approach with a flux limiter $f_e$ that is used to limit the rate at which heat can be transported (for more details, see Ref.~\onlinecite{Meezan:2020} and references therein). 
The appropriate value of $f_e$ for any given configuration is an area of active research,\cite{Farmer:2020,Farmer:2021} and we treat it as a free parameter here.
We first modeled the CH capsule N190707 using 1D \texttt{xRAGE} simulations in order to calibrate $\eta_{\rm laser}$ and $f_e$, following Ref.~\onlinecite{Ellison:2018}. 
We performed a two-dimensional scan of the parameter space, seeking to match the observed DT yield and the bang time as closely as possible. 
The best match to N190707 was found for $\eta_{\rm laser}=0.65$ and $f_e=0.05$.

Having calibrated our 1D simulations to match the existing data from N190707, we then began preliminary design simulations for the N201115 target.
As noted previously, we considered an ablator consisting of an outer pure beryllium layer and an inner boron-doped beryllium layer.
In our modeling for N201115, the \isotope[10]{B} and \isotope[9]{Be} are included as separate materials, each with their respective equations of state (EOS) from the Livermore Equations of State (LEOS) database\cite{Young:1995} and opacities as calculated from the \texttt{OPLIB} database\cite{Colgan:2016} using the \texttt{TOPS} code.\cite{Abdallah:1985} 
The materials are pre-mixed in the appropriate atomic fractions in the doped region of the ablator. 
In these initial design studies, we considered $20\,\rm at.\,\%$ \isotope[10]{B} in the inner beryllium layer; though as will be discussed subsequently, we were limited to $10\,\rm at.\,\%$ due to fabrication practicalities.
Finally, although a small fraction of \isotope[40]{Ar} is present in the ablator layers as noted previously, this was not modeled in these initial calculations.
Sensitivity to the argon concentration is considered later in Sec.~\ref{sec:sensitivity}.

\begin{figure}[t]
\includegraphics[width=\linewidth,clip=true]{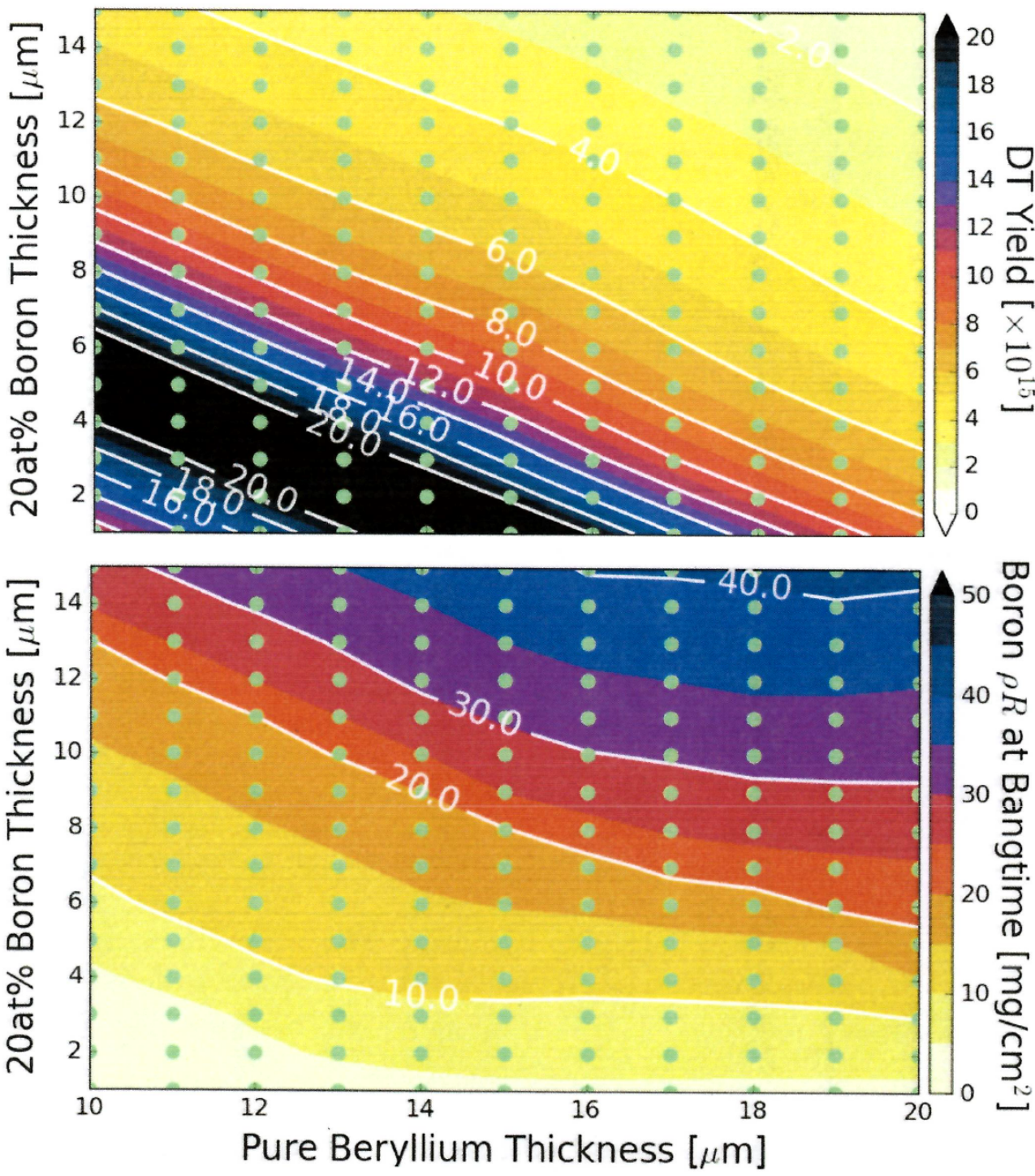}
\caption{Results of 1D \texttt{xRAGE} design optimization study with $\eta_{\rm laser}=0.65$ and $f_e=0.05$ varying the thicknesses of the pure Be outer ablator layer and the $20\,\rm at.\,\%$ \isotope[10]{B}-doped inner ablator layer. The DT neutron yield is optimized for a total ablator thickness near $15\,\rm\mu m$ (top), while the \isotope[10]{B} $\rho R$ remaining at bang time increases with thicker ablators (bottom).}
\label{fig:design_opt}
\end{figure}

We varied the thicknesses of the outer pure beryllium ablator layer and the inner boron-doped layer, and the results of this parameter scan are shown in \cref{fig:design_opt}.
The gas radius was fixed at $1485\,\rm\mu m$, and the fill was 65/35 DT at $7.86\,\rm atm$ in these calculations. 
Our goal was to maximize both the DT neutron yield and the boron $\rho R$ at stagnation in order to provide targets for the $\alpha$ particles produced during thermonuclear burn. 
In contrast to N190707, here the goal was to have a shell whose inner \isotope[10]{B}-doped portion would be mostly intact at stagnation, providing sufficient targets for the $\alpha$ particles, while the outermost pure beryllium layer was ablated off. 
We find that the DT neutron yield is optimized in this design for thin ablators, on the order of $15\,\rm\mu m$ total thickness, with a peak value just above $2\times10^{16}$ DT neutrons.
The yield is insensitive to the breakdown between pure beryllium and doped layer thickness.

The boron $\rho R$ at stagnation is increased for thicker ablator layers.
If the total ablator layer thickness is too small, the shell ends up burning through and very little boron remains near stagnation.
Thicker layers increase the boron $\rho R$, as expected, but the total yield drops considerably due to the added ablator mass.
Based on these studies, we opted for a capsule that had a $15\,\rm\mu m$ thick boron-doped beryllium inner layer and a $10\,\rm \mu m$ thick pure beryllium outer layer. 
The predicted 1D yield for such a target was $\sim4.5\times10^{15}$~DT neutrons with a boron $\rho R$ of $27.5\,\rm mg/cm^2$ at stagnation.

\subsection{Target fabrication}
\label{sec:fabrication}
During the fabrication process of the target, we conducted three different coating runs to test the addition of \isotope[10]{B} dopant to a beryllium ablator. 
The first run resulted in a leak-tight capsule with good surface quality.
However, the amount of boron introduced in the ablator was estimated to be much less than anticipated, of the order of $\sim5\,\rm at.\,\%$. 
With the second run, the \isotope[10]{B} dopant level reached $\sim20\,\rm at.\,\%$, but the target was not leak-tight. 
The third and final run produced a leak-tight capsule with an intermediate value of $\sim10\,\rm at.\,\%$ of \isotope[10]{B} dopant.
However, the beryllium ablator did not shrink as much as expected during pyrolysis of the target mandrel, and the outer radius was much larger than originally designed. 

\begin{figure}[t]
\includegraphics[width=\linewidth, trim=0.0cm 1.0cm 0.0cm 1.0cm, clip=true]{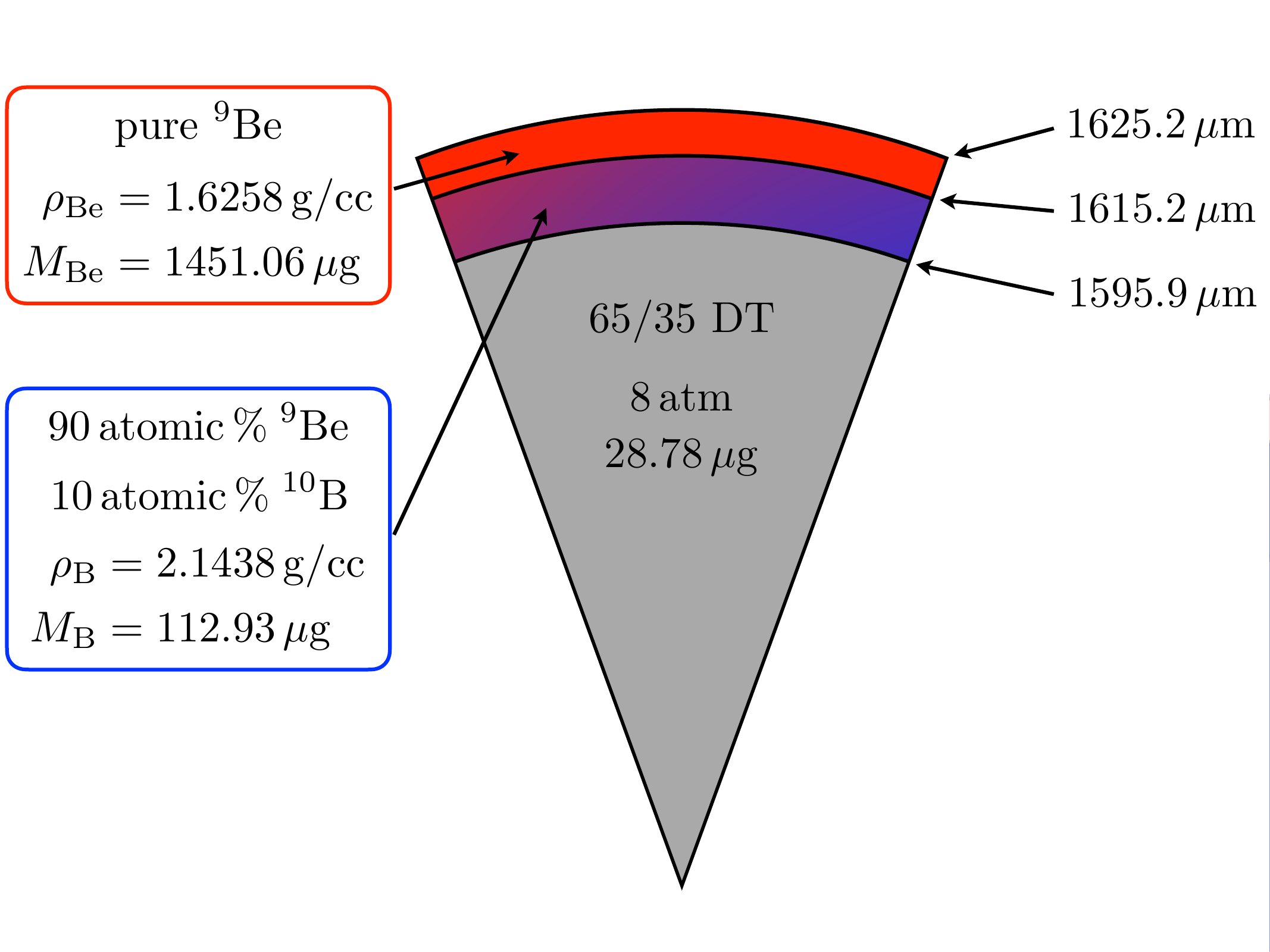}
\caption{Details of the N201115 target as fabricated by General Atomics and used in post-shot \texttt{xRAGE} calculations. 
The composite density of the ablator doped layer is computed as $\rho_{\rm total}=(f_{\rm Be}+10/9f_{\rm B})\,\rho_{\rm Be}$, where $f_{\rm X}$ is the atomic fraction of the species $\rm X$. 
The ablator also contains $0.3\,\rm at.\,\%$ of $\isotope[40]{Ar}$.}
\label{fig:design}
\end{figure}

The resulting capsule had an inner radius of $\sim 1600\,\mu\rm m$, and it was filled with $65/35\,\rm at.\,\%$ DT at $8\,\rm atm$, for a total fuel mass of $28.78\,\mu\rm g$ (\cref{fig:design}). 
There is a $\sim19\,\mu\rm m$ thick $10\,\rm at.\,\%$ boron-doped beryllium layer surrounded by a $10\,\mu\rm m$ pure beryllium layer. 
With this configuration, the total mass of the boron dopant is $112.93\,\mu\rm g$, and the average density of the doped region of the ablator was estimated at $1.64\,\rm g/cc$. 
All subsequent computational results use these target dimensions.

\section{Post-shot modeling in 1D}
\label{sec:analysis}
Although the as-built target dimensions differed considerably from the ``optimal'' design that was determined in Sec.~\ref{sec:preshot}, we proceeded with fielding the target as a proof-of-principle shot.
Here, we present our initial 1D post-shot modeling using the as-built target dimensions, we introduce the radiochemistry post-processing framework, and we perform a sensitivity analysis to the dopant concentration levels.

\subsection{\texttt{xRAGE} simulations}
\label{sec:postshot1d}
Our initial design studies neglected the small amount of argon in the ablator layers, but we now include it here by modifying the opacities for both the beryllium and boron materials.
We also modified the isotopic concentrations to include the argon component, as isotopics are important for the 3T package which impacts the laser deposition. 
However, the equations of state were left unchanged in this study.
Sensitivity to the \isotope[40]{Ar} concentration is discussed further in Sec.~\ref{sec:sensitivity}.

As mentioned previously, modeling of the CH ablator shot N190707 found very good agreement with $\eta_{\rm laser}=0.65$ and a flux limiter of $f_e=0.05$.
However, our initial 1D post-shot modeling using the as-built target dimensions for our \isotope[9]{Be}/\isotope[10]{B} ablator shot found that we could get a better match to the observed bang time (see \cref{tab:exp}) using a higher laser power multiplier, $\eta_{\rm laser}=0.85$, and the same flux limiter. This likely points to key differences in the ablation efficiencies of CH and Be. Similar behavior has been observed in \texttt{HYDRA}\citep{Marinak:2001} calculations, where a larger power multiplier is needed for Be ablators relative to CH ablators.\cite{Schmitt:pc} 
Note that this higher laser power multiplier will change the contours of boron $\rho R$ and DT yield previously identified in \cref{fig:design_opt}. The additional laser power will result in more ablation of material, so initially thicker targets are needed to compensate for this effect. The yield and boron $\rho R$ contours in \cref{fig:design_opt} will both shift up and to the right.
All the results presented in the next sections have been obtained using $\eta_{\rm laser}=0.85$ and $f_e=0.05$ in 1D \texttt{xRAGE} calculations.

\begin{figure}[t]
\includegraphics[width=\linewidth]{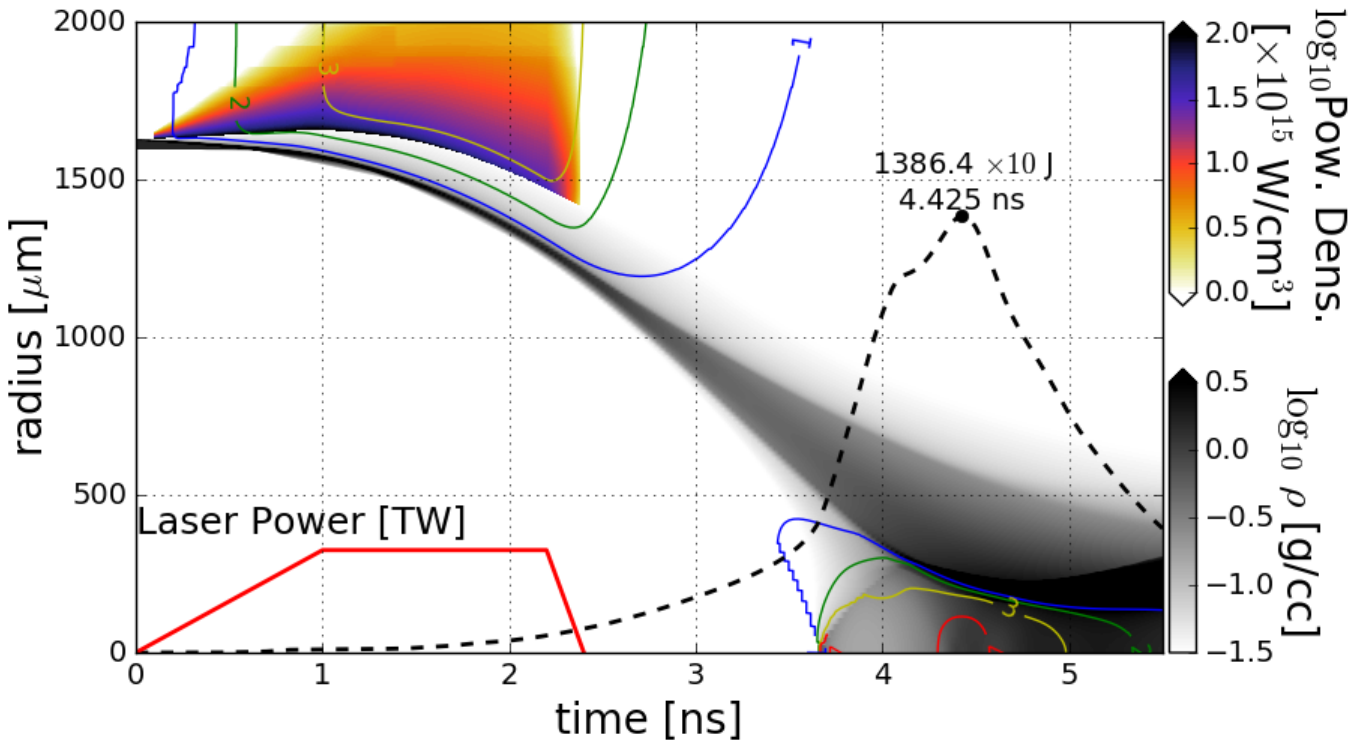}
\caption{Density contours (grayscale) and laser power deposition (color) vs $r$ and $t$ from 1D \texttt{xRAGE} simulation with $\eta_{\rm laser}=0.85$ and $f_e=0.05$ for N201115 with dimensions given in \cref{fig:design}. The laser pulse is shown as the solid red line, ion temperature contours (in keV) are shown as colored lines, and the fuel internal energy is shown as the dashed black line.}
\label{fig:1d_r-t}
\end{figure}

The dynamics of the implosion are best illustrated in the $r$--$t$ contour plot from a 1D \texttt{xRAGE} simulation, shown in \cref{fig:1d_r-t}. 
The density contours are shown in grayscale and the laser power deposition is illustrated in the color contours. 
The laser drive lasts for the first $2.4\,\rm ns$ of the implosion, ablating material off of the outer surface of the capsule and driving the remaining mass inwards. 
As the shell coasts inwards, its kinetic energy is converted into internal energy of the DT fuel, which is shown in the dashed black line of \cref{fig:1d_r-t}. 
The peak value of fuel internal energy occurs at $4.425\,\rm ns$, which nearly coincides with the capsule bang time.
Given the fuel conditions as a function of time, we calculated the DT neutron production rate, resulting in a predicted total yield of $3.31\cdot10^{15}$ DT neutrons and a burn width of $323\,\rm ps$ (see also \cref{fig:yields} and \cref{tab:1d_res}). 

\begin{figure}[t]
\includegraphics[width=\linewidth, trim=0.0cm 0.6cm 0.0cm 1.0cm, clip=true]{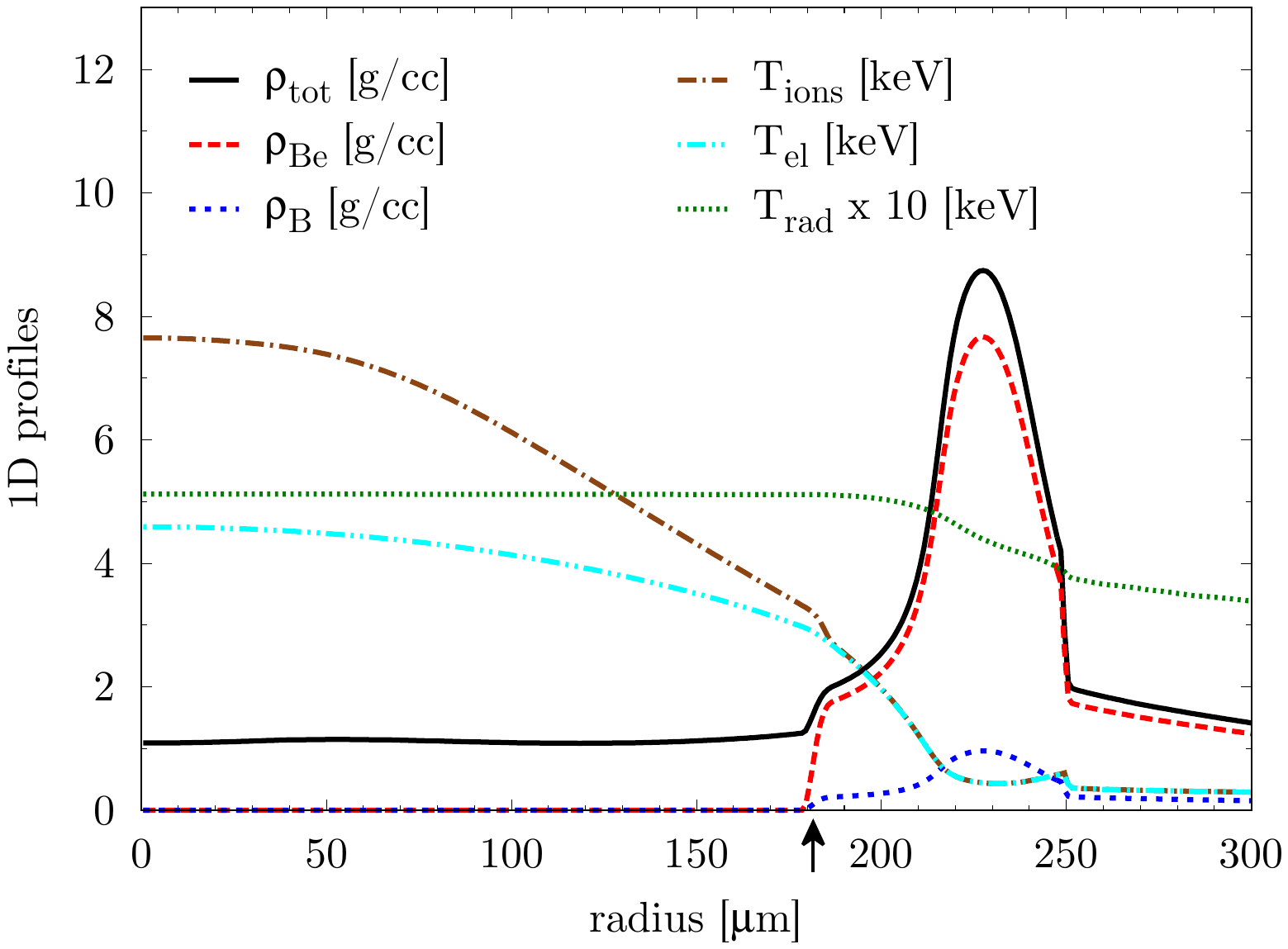}
\caption[]{Density and temperature profiles from 1D \texttt{xRAGE} calculations corresponding to the bang time conditions of \cref{fig:1d_r-t}. The position of the fuel-ablator interface at bang time is shown with a black arrow.}
\label{fig:rho-temp}
\end{figure}

The corresponding bang time density and temperature profiles (assuming no hydrodynamical mix between the fuel and the ablator shell material) are shown in \cref{fig:rho-temp}. 
At this time, the fuel-ablator interface is situated at $181.5\,\mu\rm m$, with a DT average density of $1.12\,\rm g/cc$ and a peak ablator density of $8.74\,\rm g/cc$. 
The central ion and electron temperatures are $7.65$ and $4.59\,\rm keV$, respectively. 
The radiation temperature in the capsule is much smaller, only about $512\,\rm eV$.

\subsection{RadChem analysis}
\label{sec:radchem}
We developed a charged-particle transport post-processor in order to study $\alpha$-induced reactions in the ablator material. 
Here, we describe the post-processor and apply it to our 1D post-shot simulation results.
Given the generic reaction $X(\alpha,n)Y$, the final number of nuclei $N_Y$ is given by the integral over energy and time of the $\alpha$-particle flux $\Phi_\alpha(E,t)$, multiplied by the reaction cross section $\sigma_{\alpha X}(E)$ and the initial number of nuclei $N_X$
\begin{align}
    N_Y=\int\!dt\int\!dE\,\Phi_\alpha(E,t)\,\sigma_{\alpha X}(E)\,N_X\,,
\end{align}
where the energy integral is evaluated for $\alpha$-particle energies from $3.5$ to $0\,\rm MeV$. 
In our study, $X=\isotope[10]{B},\,\isotope[9]{Be}$ and $Y=\isotope[13]{N},\,\isotope[12]{C^*}$, respectively.

The time-dependent 1D \texttt{xRAGE} density and temperature profiles have been used to track the energy loss of $\alpha$ particles produced in different regions of the fuel due to the stopping power in the target plasma (both within the DT fuel and the ablator materials). 
The $\alpha$ stopping power is calculated according to Zimmerman\cite{Zimmerman:1990} and includes contributions from free electrons and ions in the plasma. 
No bound electron contribution is considered, as the plasma is fully ionized at the temperatures considered here. 

The implosion and nuclear burn conditions of N201115 from 1D \texttt{xRAGE} computations are such that all of the $\alpha$ particles produced into the DT fuel are predicted to reach and cross the fuel-ablator interface. 
They all eventually stop within the ablator, where $\alpha$-induced reactions on \isotope[10]{B} and \isotope[9]{Be} can occur (no mix assumed). 
At bang time, $\alpha$ particles get trapped within $\sim 48\,\mu\rm m$ into the ablator that roughly corresponds to the position of the peak density of boron and beryllium (see \cref{fig:rho-temp}).

We note that, as a result of the relatively large size of the fuel region, $\alpha$ particles loose a non-negligible amount of energy before reaching the ablator ($\sim0.87\,\rm MeV$ on average at bang time).
This greatly affects the yield of $\alpha$-induced reactions as the relevant cross sections are strongly energy dependent (see \cref{fig:cross}).
The $14.1\,\rm MeV$ DT neutrons are in stark contrast to this, as they can travel almost unperturbed through the plasma (minimal energy loss), producing \isotope[41]{Ar} in the ablator via the reaction $\isotope[40]{Ar}(n,\gamma)\isotope[41]{Ar}$, that has a cross section of $\sigma(14.1\,{\rm MeV})=0.449\,\rm mb$.\cite{Otuka:2014}

\begin{figure*}[t]
\includegraphics[width=0.27\linewidth,trim=3.85cm 0.90cm 7.4cm 1.5cm, clip=true]{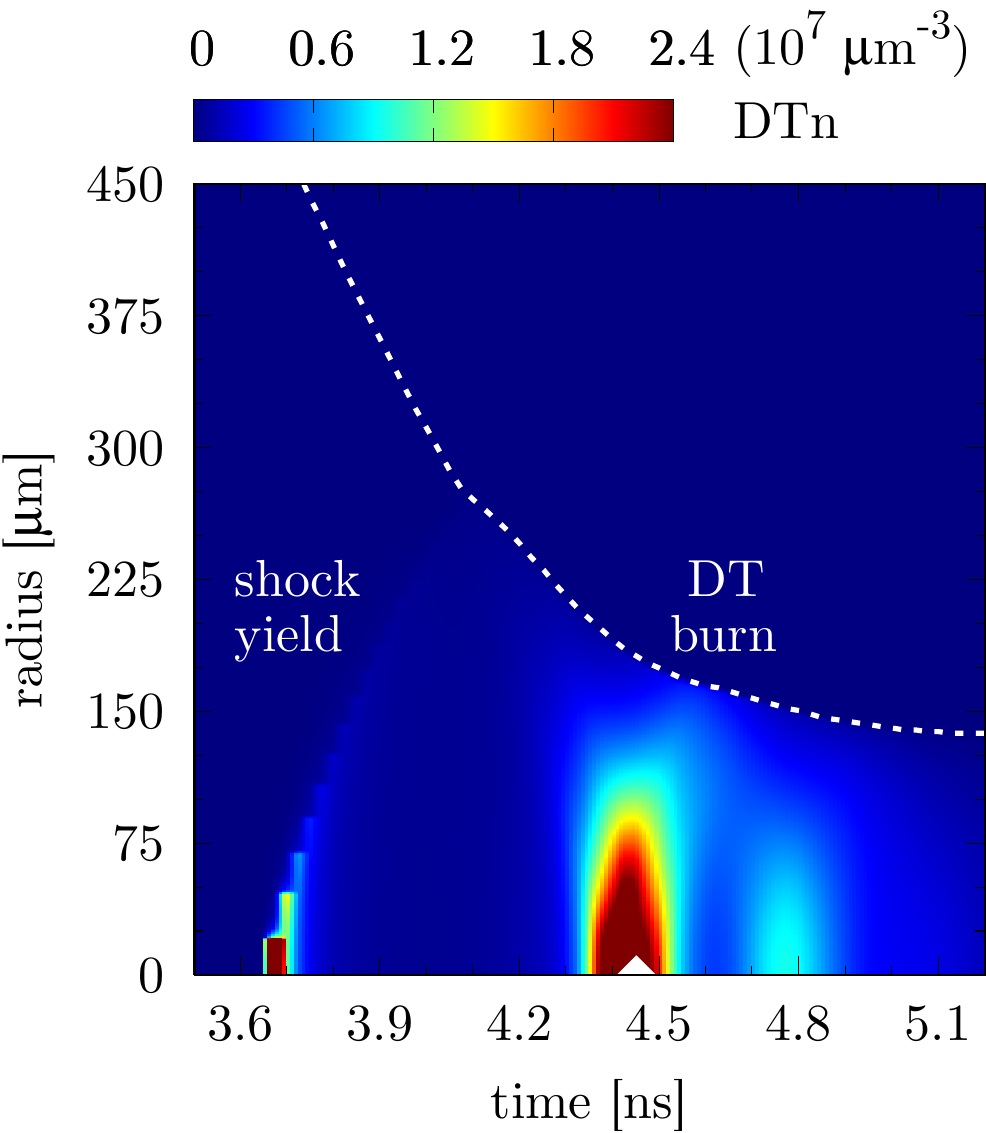}\hfill
\includegraphics[width=0.22\linewidth,trim=7.05cm 0.90cm 7.4cm 1.5cm, clip=true]{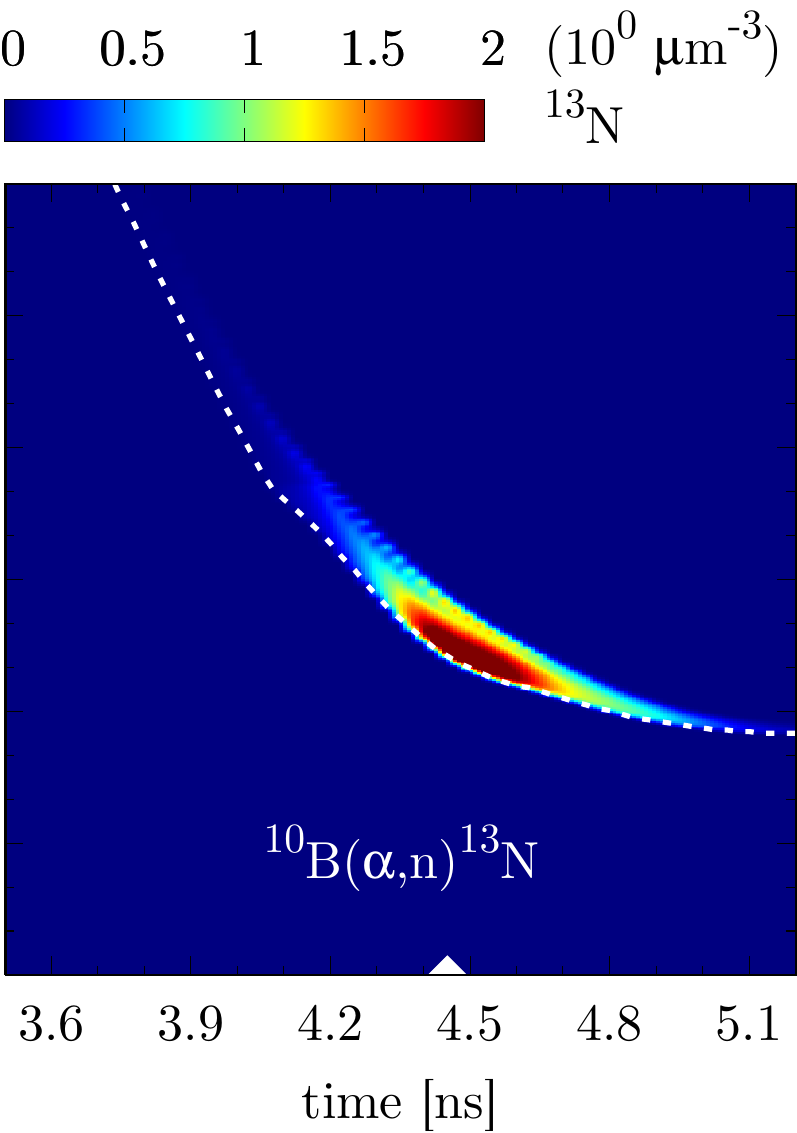}\hfill
\includegraphics[width=0.22\linewidth,trim=7.05cm 0.90cm 7.4cm 1.5cm, clip=true]{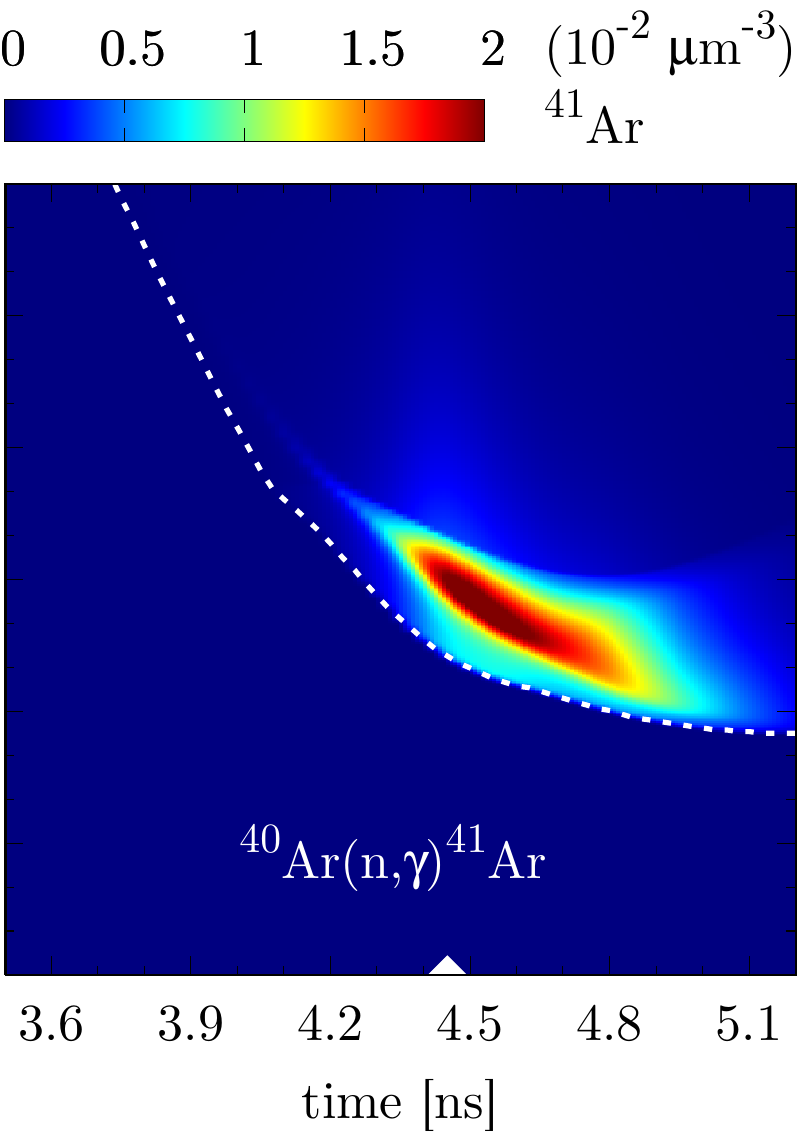}\hfill
\includegraphics[width=0.22\linewidth,trim=7.05cm 0.90cm 7.4cm 1.5cm, clip=true]{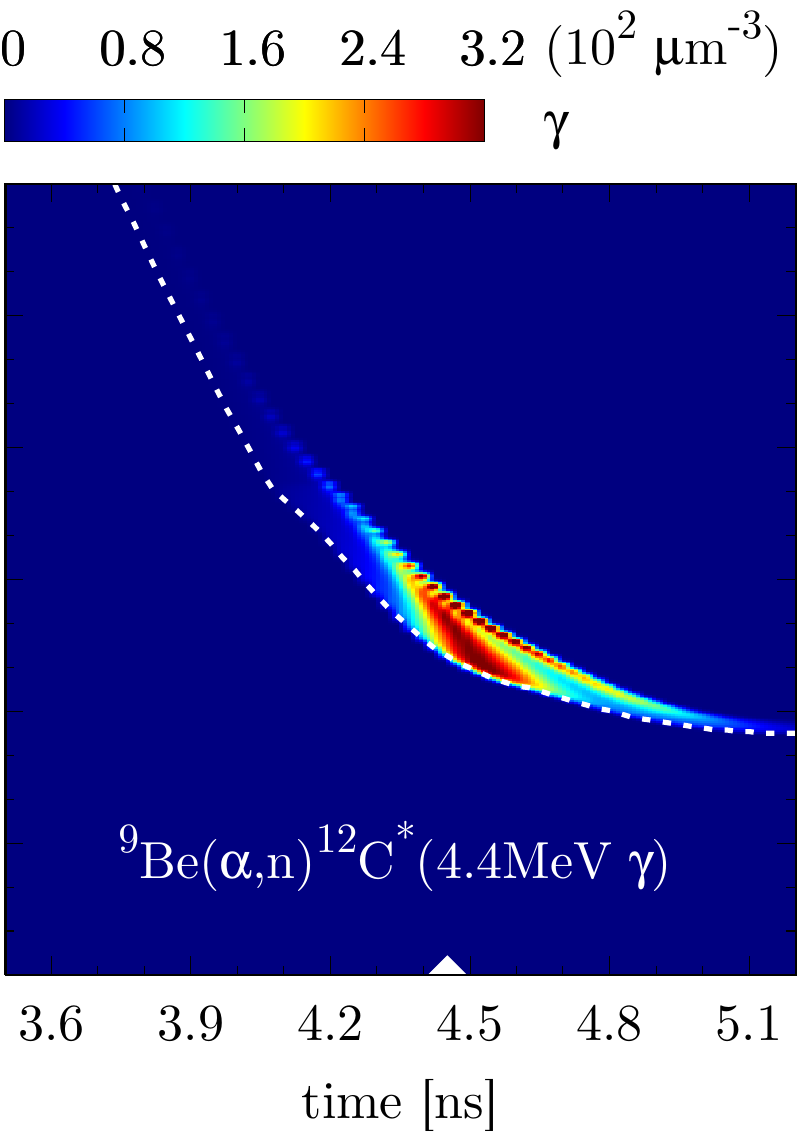}
\caption[]{$r-t$ contour plot of the implosion of the N201115 target. From left to right, particle yield per unit volume for DT$n$, \isotope[13]{N}, \isotope[41]{Ar}, and $4.4\,\rm MeV$ $\gamma$ from the RadChem analysis of 1D \texttt{xRAGE} computations. The white dashed curve corresponds to the position of the fuel-ablator interface. The white upward triangle indicates bang time, $4.450\,\rm ns$. Imperfections in the color map are due to the somewhat limited temporal resolution $(25\,\rm ps)$ of the \texttt{xRAGE} outputs.}
\label{fig:yields_2d}
\end{figure*}

\begin{figure}[b]
\includegraphics[width=\linewidth, trim=0.0cm 0.6cm 0.0cm 1.0cm, clip=true]{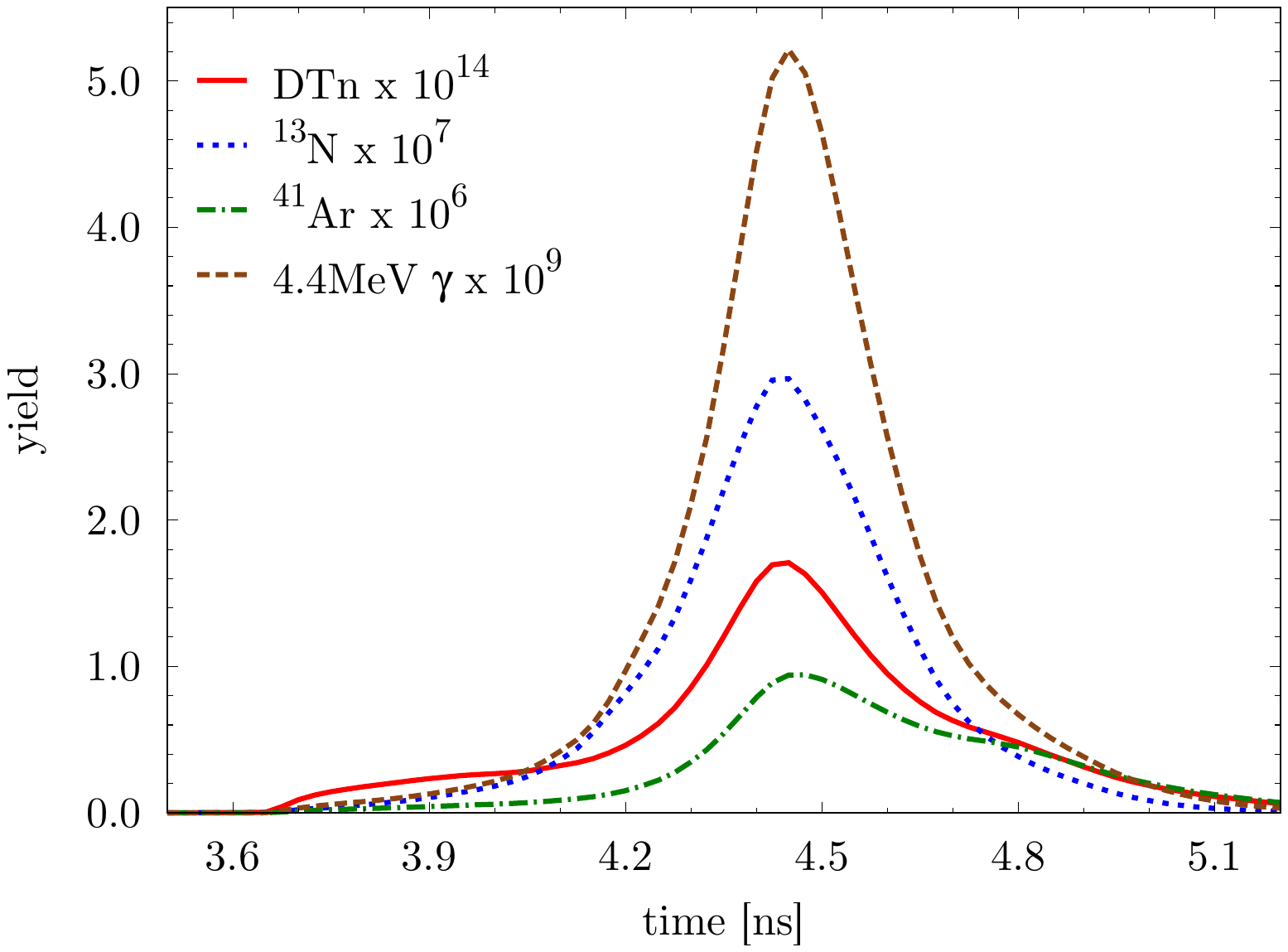}
\caption[]{DT$n$, \isotope[13]{N}, \isotope[41]{Ar}, and $4.4\,\rm MeV$ $\gamma$ time-dependent yields.}
\label{fig:yields}
\end{figure}

\Cref{fig:yields_2d} shows the $r-t$ contour plot for the predicted particle yields per unit volume achieved during the N201115 implosion: left to right for DT$n$, \isotope[13]{N}, \isotope[41]{Ar}, and $4.4\,\rm MeV$ $\gamma$. 
Bang time is highlighted with a white upward triangle, and the position of the fuel-ablator interface is also displayed with a white dashed curve. 
The latter matches the contours for DT neutrons and $\alpha/n$-induced reactions as no hydrodynamical mix between the fuel and the ablator material has been considered in this run. 
The DT$n$ profile shows an early neutron production for $t\lesssim4.1\,\rm ns$. 
This corresponds to the shock yield and has minimal impact on $\alpha/n$-induced reactions in the ablator. 
During the following DT nuclear burn, the fuel region is compressed from $\sim275\,\mu\rm m$ to $\sim150\,\mu\rm m$, with the peak neutron density production at bang time localized within $\sim100\,\mu\rm m$ from the center of the capsule. 
The $\alpha$-induced reactions on \isotope[10]{B} and \isotope[9]{Be} are confined in a narrow region close to the fuel-ablator interface, as the alpha particles quickly lose energy into the ablator material and eventually stop. 
The \isotope[41]{Ar} production covers instead a wider spatial region as the DT neutrons have minimal energy loss traveling through the plasma, and they can, thus, interact with the entire ablator layer before escaping the capsule.
The time-dependent (position-integrated) yields are shown in \cref{fig:yields}. 
The DT$n$ shock yield is identifiable for earlier times $(t\lesssim 4.1\,\rm ns)$. 
As already mentioned, this earlier signal does not significantly contribute to $\alpha/n$-induced reactions, as also visible in this figure. 

A summary of the 1D \texttt{xRAGE}-RadChem analysis is reported in \cref{tab:1d_res}. 
The predicted DT$n$ yield is $3.31\cdot10^{15}$. 
The expected \isotope[13]{N} and \isotope[41]{Ar} signals are of the order of $10^8$ and $10^7$, respectively. 
Both signals exceed the corresponding RAGS thresholds ($4.65\cdot10^8>Y_{\rm{th}}(\isotope[13]{N})=10^5$ and $1.94\cdot10^7>Y_{\rm{th}}(\isotope[41]{Ar})=10^5$), and are, thus, measurable. 
The $4.4\,\rm MeV$ $\gamma$ signal from the $\isotope[9]{Be}(\alpha,n)\isotope[12]{C^*}$ reaction is also expected to be above threshold $(7.30\cdot10^{10}>Y_{\rm th}(4.4\,\rm MeV\,\gamma)=4\cdot10^{10})$, and, thus, theoretically measurable by the GRH detector.

The combination of these radiochemical signals, in particular their ratios, can be used to assess mixing. As pointed out in Ref.~\onlinecite{Colvin2008}, in mix calculations of ICF capsules, ratios of thermonuclear yield and radiochemical signals are sensitive to mix. In particular, different reactions lead to different yield ratios when mix is involved, and this can be used to indirectly assess the level of mixing. $\alpha$- and $n$-induced reactions are differently affected by mix, the former manifesting larger variations due to non-trivial changes in the $\alpha$ stopping power in the fuel when mix occurs. The combined analysis of \isotope[13]{N} and \isotope[41]{Ar} signals is, thus, a valuable tool for mix diagnostic. The $4.4\,\rm MeV\;\gamma$ signal can potentially be used as a complementary diagnostic, as it is still largely affected by mix ($\alpha$-induced reactions), but the measurement is done with an independent detector. Preliminary 1D \texttt{xRAGE} calculations including BHR mixing\cite{Besnard:1989}, show, indeed, $4-6\%$ variations in the RadChem ratios for modest mixing scenarios, not commensurate with changes in the DT$n$ yield. A detailed discussion of RadChem mix diagnostic for boron/beryllium ablator capsules goes beyond the scope of this work as the current target was not designed for mix sensitivity, but rather to test our ability to measure the $^{10}{\rm B}(\alpha,n)^{13}{\rm N}$ reaction. Modifying the current design to be more sensitive to mix is the subject of a future investigation.

\begin{table}[h]
\caption{Summary of the 1D \texttt{xRAGE}-RadChem analysis of the design shown in \cref{fig:design}.}
\label{tab:1d_res}
\begin{ruledtabular}
\begin{tabular}{lc}
\isotope[10]{B} dopant level    & $10\,\rm at.\,\%$         \\
\isotope[40]{Ar} concentration  & $0.3\,\rm at.\,\%$        \\
Bang time                       & $4.450\,\rm ns$              \\
Burn width                      & $323\,\rm ps$                \\
DT$n$ yield                     & $3.31\cdot10^{15}$           \\
\isotope[13]{N} yield           & $4.65\cdot10^{8\phantom{0}}$ \\
\isotope[41]{Ar} yield          & $1.94\cdot10^{7\phantom{0}}$ \\
$\gamma$ $(4.4\,\rm MeV)$ yield & $7.30\cdot10^{10}$           \\
\end{tabular}
\end{ruledtabular}
\end{table}

\subsection{Sensitivity study}
\label{sec:sensitivity}
The \texttt{xRAGE}-RadChem analysis presented so far refers to the capsule design illustrated in \cref{fig:design} that includes a $10\,\rm at.\,\%$ \isotope[10]{B} dopant level and a $0.3\,\rm at.\,\%$ \isotope[40]{Ar} concentration in the ablator. 
An accurate determination of the boron and argon concentrations in the manufactured N201115-001 capsule is, however, challenging. 
In order to assess the sensitivity of our simulations to variations in the boron and argon concentrations, we ran few extra scenarios.

We first changed the \isotope[10]{B} dopant level up to twice the design value (up to $20\,\rm at.\,\%$).
This does not significantly modify the implosion dynamics.
Bang time remains unchanged, and ion, electron, and radiation temperatures are subject to minimal variations throughout the implosion.
The total DT$n$ yield decreases by at most $\sim6\%$, implying a similar change to the \isotope[41]{Ar} signal.
Changes to the nitrogen (gamma) yield are linearly proportional to the increased (decreased) concentration of boron (beryllium), as one might have naively expected.
A $25\%$ uncertainty on the boron concentration will imply a $\sim20\%$ uncertainty on the predicted $\alpha$-induced reaction yields, with almost no change to the other relevant physical quantities.

Changes in the argon concentration, instead, are less trivial and result in larger modifications of the implosion and burn properties (see \cref{tab:40ar_conc}).
Adding argon results in the ablator becoming more diffuse during the coasting phase, likely an effect of the increased opacity. 
Essentially, the argon acts as a sink for photons, and the ablator heats up and expands during the coasting as a result of absorption of this ``extra'' energy.
During the laser drive phase, the peak radiation temperature in the boron-dopant material is $96\,\rm eV$ without any argon, while with $1.2\,\rm at.\,\%$ \isotope[40]{Ar} the peak radiation temperature in the boron is $121\,\rm eV$. 
There is not a large effect on the inner surface trajectory, and the reported laser energy absorbed by the capsule is comparable in all cases. 
Increasing the \isotope[40]{Ar} concentration results in a lower peak ablator density at stagnation, owing to the more diffuse ablator. 
Increasing the argon percentage drops the DT$n$ yield considerably and results in larger burn widths. 
This has a considerable impact on $\alpha$-induced reactions, with \isotope[13]{N} and $4.4\,\rm MeV$ $\gamma$ yields decreasing by roughly a factor of 4 moving from no argon to $1.2\,\rm at.\,\%$ \isotope[40]{Ar} concentration.

From the above analysis, it is evident that argon is not a passive ride-along for direct-drive implosions. 
It results in reduced performance, owing to a more diffuse ablator during the coasting phase. 
However, it is possible that the detrimental effects of argon in the ablator might be mitigated somewhat by a shorter coast time, giving the ablator less time to decompress as a result of the absorbed radiation energy.

\begin{table}[h]
\caption{Summary of the 1D \texttt{xRAGE}-RadChem analysis for different \isotope[40]{Ar} concentrations. For these runs, the \isotope[10]{B} dopant level has been set to $10\,\rm at.\,\%$.}
\label{tab:40ar_conc}
\begin{ruledtabular}
\begin{tabular}{lcccc}
\isotope[40]{Ar} conc. & $0\,\rm at.\,\%$ & $0.3\,\rm at.\,\%$ & $0.6\,\rm at.\,\%$ & $1.2\,\rm at.\,\%$ \\
\midrule
Bang time                       & $4.450\,\rm ns$              & $4.450\,\rm ns$              & $4.425\,\rm ns$              & $4.425\,\rm ns$              \\
Burn width                      & $328\,\rm ps$                & $323\,\rm ps$                & $332\,\rm ps$                & $365\,\rm ps$                \\
DT$n$ yield                     & $4.50\cdot10^{15}$           & $3.31\cdot10^{15}$           & $2.77\cdot10^{15}$           & $2.19\cdot10^{15}$           \\
\isotope[13]{N} yield           & $8.00\cdot10^{8\phantom{0}}$ & $4.65\cdot10^{8\phantom{0}}$ & $3.31\cdot10^{8\phantom{0}}$ & $2.00\cdot10^{8\phantom{0}}$ \\
\isotope[41]{Ar} yield          & $0$                          & $1.94\cdot10^{7\phantom{0}}$ & $3.07\cdot10^{7\phantom{0}}$ & $4.47\cdot10^{7\phantom{0}}$ \\
$\gamma$ $(4.4\,\rm MeV)$       & $1.26\cdot10^{11}$           & $7.30\cdot10^{10}$           & $5.13\cdot10^{10}$           & $3.18\cdot10^{10}$           \\
\end{tabular}
\end{ruledtabular}
\end{table}

\section{Experimental results}
\label{sec:exp}
The NIF shot N201115-001 produced an estimated $\sim4\cdot10^{14}$ DT neutrons, only $\sim12\%$ of the predicted 1D yield, with significant shape asymmetry. 
The capsule bang time occurred at $4.83\pm0.13\,\rm ns$ $(4.5\pm0.5\,\rm ns)$ based on the SPIDER (GRH) measurement, and the burn duration (full-width-at-half-maximum) was $600\pm75\,\rm ps$ according to the GRH analysis (see \cref{tab:exp} for details).

\begin{figure*}[t]
\begin{minipage}[left]{0.495\linewidth}
\includegraphics[width=\linewidth, trim=0cm -2.7cm 0cm -2.7cm, clip]{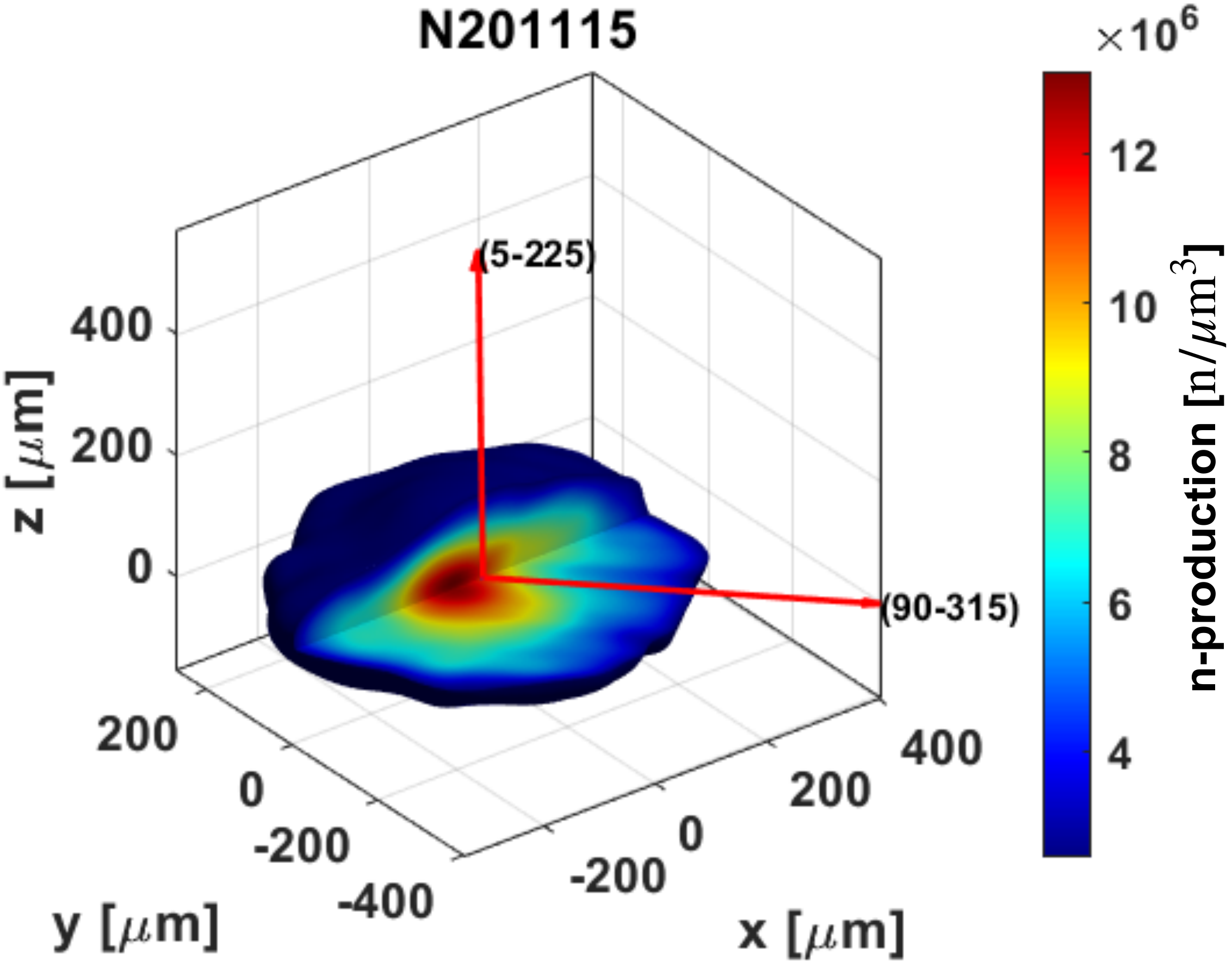} 
\begin{ruledtabular}
\begin{tabular}{cc}
\multicolumn{2}{c}{Legendre decomposition} \\
\midrule
Polar view & Equatorial view \\
$m_0=299.14\pm2.13\,\mu\rm m$ & $P_0=211.95\pm2.54\,\mu\rm m$ \\
$m_2/m_0=5.97\pm0.82\%$       & $P_2/P_0=-48.50\pm0.61\%$     \\
$m_6/m_0=3.14\pm1.24\%$       & $P_4/P_0=17.41\pm0.57\%$      \\
\end{tabular}
\end{ruledtabular}
\end{minipage}
\hfill
\begin{minipage}[right]{0.495\linewidth}
\hfill\includegraphics[width=0.96\linewidth]{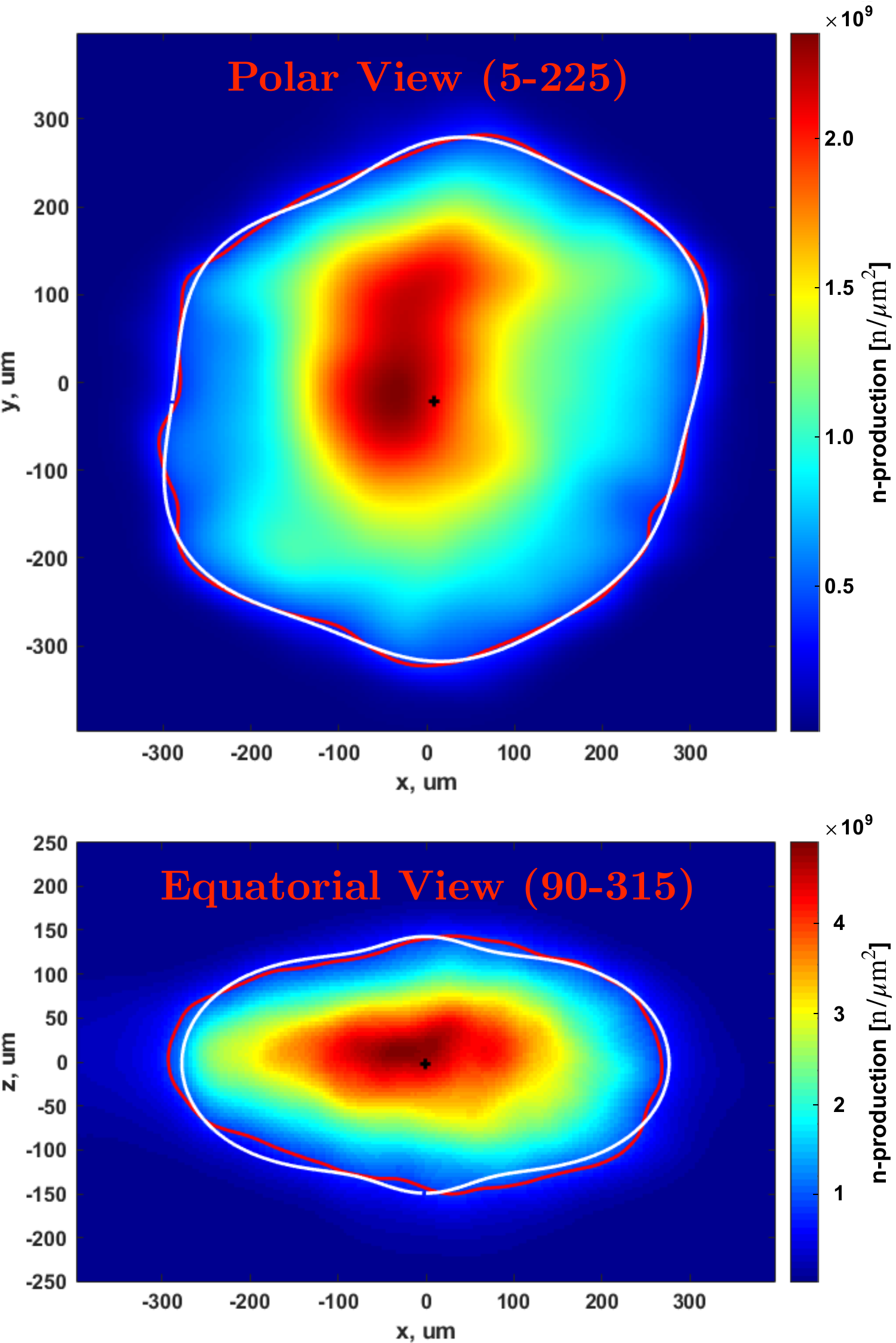}
\end{minipage}
\caption[]{Reconstructed DT neutron production distribution of the N201115-001 PDXP shot. Top left panel: full 3D reconstruction. Right panels: polar view (top) and equatorial view (bottom). 
The red line is the 17\% contour relative to the maximum intensity. The white line is the result of the Legendre polynomial fit to the 17\% contour.
Bottom left table: dominant components of the Legendre decomposition of the polar and equatorial views. Uncertainties include both statistical and systematic errors added in quadrature.}
\label{fig:shot_3d}
\end{figure*}

The neutron production region was very large, with a sizable Legendre mode $P_2$ asymmetry and modest azimuthal asymmetry. 
A 3D reconstruction of the neutron production region for this shot is shown in \cref{fig:shot_3d}, produced from two independent lines-of-sight utilizing the procedure described in Ref.~\onlinecite{Volegov:2021}. 
The first line of sight views the capsule from near the northern pole of the target chamber at $\theta=5^\circ$ and $\phi=225^\circ$, while the second line of sight views the capsule from the equator looking from $\theta=90^\circ$ and $\phi=315^\circ$. 
The 2D images from each diagnostic port are also shown in \cref{fig:shot_3d}.

The dominant Fourier component from a decomposition of the polar view is $m=2$ with $m_2/m_0\sim5.97\%$, and there was also a significant $m=6$ azimuthal feature with $m_6/m_0\sim3.14\%$. 
The azimuthal asymmetry is here attributed to the fact that two NIF quads (sets of four beams) had to be dropped for this shot. 
However, these dropped beams do not explain the significantly pancaked (oblate) implosion, which was found to have very large Legendre $P_2$ and $P_4$ components as extracted from the equatorial view. 
The average $(P_0)$ radius of the neutron production region was estimated to be quite large, $\sim212\,\mu\rm m$, with $P_2/P_0\sim-48.5\%$ and $P_4/P_0\sim17.4\%$. 
These numbers are summarized in the inset table of \cref{fig:shot_3d}. 
As will be demonstrated subsequently, this shape asymmetry can be reasonably reproduced in preliminary 2D \texttt{xRAGE} post-shot modeling.

\subsection{Radiochemical signatures}
Both \isotope[13]{N} and \isotope[41]{Ar} were detected at RAGS. 
The \isotope[13]{N} is collected directly into the RAGS abort tank, bypassing the getters, and assayed by measuring its $t_{1/2}=9.97\,\rm min$ decay by positron emission. 
The RAGS collection efficiency of \isotope[13]{N} is not fully characterized at this time. 
The collection efficiency has been measured for those circumstances where \isotope[13]{N} is produced in a \textit{Hohlraum} via the $\isotope[14]{N}(n,2n)\isotope[13]{N}$ reaction, but this is not necessarily the same as the collection efficiency for direct drive shots because possible nitrogen molecular formation in the NIF chamber could depend on configuration geometry. 
Thus, although we saw a very clear \isotope[13]{N} signal at RAGS, we were only able to estimate the number of \isotope[13]{N} atoms produced at the time of capsule burn.
The quoted uncertainty on the \isotope[13]{N} signal (see \cref{tab:exp}) is conservatively estimated at the $20\%$ level. This is derived from the estimated detector efficiencies based on previous measurements and the statistics on the decay signal.\cite{Cassata:pc} The uncertainty stemming from the error in the \isotope[10]{B} concentration in the manufactured capsule cannot be estimated at this time, due to the difficulties in the accurate determination of the boron dopant level from General Atomics.

The collection scheme for \isotope[41]{Ar} at RAGS is described in Ref.~\onlinecite{Wilson:2017}, and it is assayed using a calibrated HPGe detector to measure the $\gamma$-rays emitted in $t_{1/2}=109.61\,\rm min$ beta-decay of \isotope[41]{Ar}. 
The analysis of the measured \isotope[41]{Ar} signal is more delicate compared to \isotope[13]{N}. One would need more than $10^5$ argon atoms to quantify the answer with a 12\% uncertainty.\cite{Cassata:pc} At and below the $10^5$ threshold, we can determine that the signal is \isotope[41]{Ar} from its half-life but cannot put an accurate figure (with a quotable uncertainty) on the number of atoms collected. Because of this and the uncertainty in the \isotope[40]{Ar} concentration in the manufactured capsule, inferring the areal density of the remaining ablator material at the time of the DT burn from the \isotope[41]{Ar} signal~\cite{Wilson:2017} would not be accurate, and it has not been attempted.

The measured signals for both \isotope[13]{N} and \isotope[41]{Ar} were considerably lower than expected based on the 1D pre-shot predictions, and at least part of this is ascribed to the reduced DT$n$ yield. 
Note that the $^{13}$N yield is lower than expected pre-shot calculations by more than the reduced DT$n$ yield. First, the exact amount of \isotope[10]{B} included in the doped ablator region was not able to be well characterized by General Atomics, so there is uncertainty in the initial amount present in the target (see Sec.~\ref{sec:sensitivity}). Second, the asymmetric nature of the implosion plays a crucial role. The asymmetric distribution of the fuel and ablator materials strongly affects the $\alpha$ particle transport in the plasma. The $\alpha$ stopping power is subject to non-trivial changes that clearly reflect in the RadChem signals due to the strongly energy-dependent cross sections of the relevant reactions.

The signal from the GRH diagnostic was low and quite noisy, complicating the analysis. 
The $4\cdot10^{10}$ threshold quoted in \cref{tab:exp} is the required threshold for a measurement to make a detectable $4.4\,\rm MeV$ $\gamma$ signal independent of the DT yield. 
For a nominal signal made up of both $4.4\,\rm MeV$ $\gamma$ rays and the DT fusion $\gamma$ rays, a $>4\cdot10^{10}$ $\gamma$ ray contribution should produce a detectable change ($>10\%$ relative increase of the signal). Since the gain settings for the GRH detector were set for a higher signal than actually achieved, we effectively obtained a near null signal. However, we can still provide a conservative upper limit for the signal (see \cref{tab:exp}). If $2\cdot10^{12}$ $4.4\,\rm MeV$ $\gamma$ rays were actually released, we would have seen a clear physics signal. Since we did not, we can estimate that the number of $4.4\,\rm MeV$ $\gamma$ rays was below that upper limit.
Note that, because of the overall noisy $\gamma$ signal, the GRH estimate of the burn width is affected by a relatively large uncertainty (\cref{tab:exp}). For cleaner data, the uncertainty is typically $30$ to $15\,\rm ps$.

\subsection{Post-shot modeling in 2D}
In order to better understand the source of the shape asymmetry, we performed a preliminary 2D post-shot computation. 
The as-fired laser beam configuration, including the beam pointings and elliptical beam spots, was imported into \texttt{xRAGE}, allowing us to capture the polar aspects of the drive though the azimuthal variation was artificially symmetrized here. 
In our initial computation, we used the as-designed laser pulse with all beams at nominally the same power. 
Further, we did not drop the missing quads here, which would require 3D computations to fully capture; a 3D laser ray-trace capability is being developed for \texttt{xRAGE}, but it is not yet available. 
Further, the experiment used wavelength detuning on some of the beams in an attempt to mitigate CBET. 
While a CBET package is available in \texttt{xRAGE}, it increases the computational cost considerably, and so, this wavelength detuning is also neglected here.
Finally, our initial 2D simulation neglected the argon dopant in the ablator material. 
All of these effects will be considered more fully in future simulations.

\begin{figure*}[t]
\includegraphics{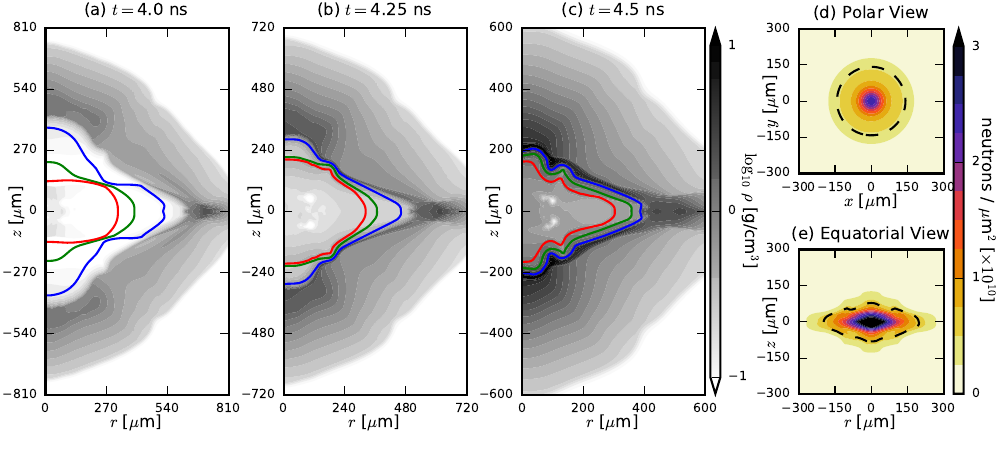}
\caption[]{Log of mass density (grayscale contours) and ion temperature contours ($\rm blue=1\,keV$, $\rm green=2\,keV$, $\rm red=3\,keV$) from a 2D \texttt{xRAGE} computation of N201115 at (a)~$4.0\,\rm ns$, (b)~$4.25\,\rm ns$, and (c)~$4.5\,\rm ns$. Synthetic neutron pinhole images integrated over all time looking at the capsule from (d)~the pole and (e)~the equator. The dashed black contours indicate 17\% of the maximum value.}
\label{fig:2d-rho}
\end{figure*}

Results from this preliminary 2D simulation are shown in \cref{fig:2d-rho}. 
Mass density contours are shown in \cref{fig:2d-rho}(a)-\ref{fig:2d-rho}(c) in grayscale at several points leading up to stagnation, and ion temperature contours are also shown as colored lines: $\rm blue=1$~keV, $\rm green=2$~keV, and $\rm red=3$~keV.
It is immediately clear that both the incoming unablated material and the central gas region are very oblate as a result of the polar laser drive configuration. 
The 2D simulation results were processed to estimate the neutron production rate as a function of time, and the total neutron production rate was integrated in time for all cells.
Bang time for this simulation occurs at $4.55\,\rm ns$, which is close to the SPIDER result and consistent with the GRH estimate (see~\cref{tab:exp}). 
The simulated DT neutron yield is $8.456\cdot10^{14}$, roughly 26\% of the predicted 1D neutron yield, but only a factor of $\sim2.1\times$ higher than the experimentally measured value. 
It is expected that the simulated yield will be further reduced when azimuthal asymmetries from the dropped quads and engineering defects, such as the capsule fill tube, are also incorporated into the simulation.
These values are summarized in \cref{tab:2d_res}.

The total neutron production rate integrated in time for all cells was used as the source in computing line-integrated neutron emission data for pinholes placed at the pole and equator of the capsule, enabling more direct comparison to the experimentally reported neutron images.
Synthetic neutron pinhole images are shown in \cref{fig:2d-rho}(d) and \ref{fig:2d-rho}(e) for a polar and equatorial view, respectively.
The dashed black contours highlight 17\% of the maximum value in each image. 
For the polar view, the 17\% contour has a radius of $142.0\,\rm \mu m$, which is only about half of the reported $m=0$ radius of $299.14\pm2.13\,\rm \mu m$.
Legendre analysis of the 17\% contour in the equatorial view reveals that $P_0=120.7\,\rm \mu m$, $P_2/P_0=-66.5\%$, and $P_4/P_0=37.4\%$. 
These values are also summarized in \cref{tab:2d_res}.
Recall that the experimentally measured values
were $P_0=211.95\ \mu$m, $P_2/P_0=-48.50\%$ and $P_4/P_0=17.41\%$. 

Although our preliminary 2D \texttt{xRAGE} calculation indicates a smaller burning region compared to what can be inferred from the neutron images, the neutron production region has a distinct oblate shape which is qualitatively consistent with the experimental results.
One possible explanation for the difference is the length of the burn width: the simulation burn width is $350\,\rm ps$ compared to the experimental burn width of $\sim600\,\rm ps$.
The longer burn width in the experiment corresponds to neutrons being produced away from the time of peak convergence.
Since these neutron images are time-integrated, this would result in the experimental neutron production region becoming larger in size, relative to simulation.
Further, as noted previously, our 2D \texttt{xRAGE} calculation is also missing some elements that can affect both DT yield and implosion shape. 
In spite of this, however, our current results are encouraging, and we expect that the agreement with experimental results will be improved as our modeling fidelity increases.

\begin{table}[h]
\caption{Summary of the 2D \texttt{xRAGE} results. The Fourier and Legendre decompositions of the 17\% contours relative to the maximum intensity are shown for the polar and equatorial views, respectively.}
\label{tab:2d_res}
\begin{ruledtabular}
\begin{tabular}{cc}
\isotope[10]{B} dopant level   & $10\,\rm at.\,\%$  \\
\isotope[40]{Ar} concentration & $0\,\rm at.\,\%$ \\
Bang time                      & $4.550\,\rm ns$  \\
Burn width                     & $350\,\rm ps$       \\
DT$n$ yield                    & $8.456\cdot10^{14}$  \\
\midrule
Polar view             & Equatorial view        \\
$m_0=142.0\,\mu\rm m$  & $P_0=120.7\,\mu\rm m$    \\
                       & $P_2/P_0=-66.5\%$     \\
                       & $P_4/P_0=37.4\%$      \\
\end{tabular}
\end{ruledtabular}
\end{table}

\subsection{Discussion}
The significant asymmetry observed in both the experimental data for N201115 and the preliminary post-shot \texttt{xRAGE} simulations clearly demands a more sophisticated post-processing of the radiochemical signatures beyond the simple 1D model introduced previously. 
Work is ongoing to develop such a framework, but a detailed comparison of the observed radiochemical measurements to our simulation predictions is not available at this time.
It is also expected that hydrodynamic instabilities, which lead to the mixing of the boron/beryllium ablator material into the DT fuel region, can alter the expected RadChem signals. 
Disentangling these effects is the subject of ongoing investigations.

One point worth mentioning is that, as noted previously, this capsule was quite a bit larger, $\sim1625\,\mu\rm m$ outer radius, than the original design dimensions based on N190707-001, which was only $\sim1480\,\mu\rm m$ outer radius. 
We did not attempt to alter the beam pointings to account for the larger radius capsule of N201115-001, so this likely accounts for at least some of the significant asymmetry observed here. 
However, it should be noted that N190707-001 also had a somewhat oblate neutron emission region\citep{Yeamans:2021}, although the performance was considerably greater ($4.81\cdot10^{15}$ DT neutrons). 
The greater performance of N190707 is expected since our capsule had considerably more ablator mass by design in order to leave some \isotope[10]{B} near the fuel region.
In order to get an estimate of how much the larger radius capsule changed the shape, we ran a preliminary 2D \texttt{xRAGE} simulation of N190707 using a lower laser power multiplier of $\eta_\text{laser}=0.65$ to approximately match bang time for this CH ablator capsule. 
However, we still neglected CBET and the wavelength detuning in this first-pass analysis. 
We found that even the smaller radius of N190707 still produced a fairly oblate implosion in \texttt{xRAGE}. 
That is, even had the boron/beryllium ablator capsule been the correct radius, we would likely still have observed an oblate implosion with reduced performance.
Clearly, additional work is needed to improve the drive conditions for these polar-drive targets.

\section{Conclusions}
\label{sec:concl}
The goal of the current experiment was to test the feasibility of producing \isotope[13]{N} through the $^{10}\text{B}(\alpha,n)^{13}\text{N}$ reaction in a PDXP at the NIF. 
This was motivated by the need to develop an alpha-particle-induced RadChem mix diagnostic for more complicated inertial confinement fusion capsule designs. 
We observed a very robust \isotope[13]{N} signal using the RAGS facility at NIF despite the fact that our PDXP yield was quite suppressed compared to 1D clean estimates. 
Our preliminary post-shot computations in \texttt{xRAGE} indicate that the oblate shape contributes to some of the reduced performance: the 2D ``clean'' yield is 26\% of the 1D clean yield. 
However, this by itself does not explain all of the degradation since the 2D yield is still $\sim2\times$ larger than the experimentally observed value.
The significant asymmetry also complicates the interpretation of the radiochemical signals, and work is ongoing to extend the radiochemistry analysis to higher dimensions.
Further optimization of the laser beam configuration for these targets is also under way. 
However, the observed \isotope[13]{N} signal strongly suggests that alpha-particle-induced radiochemistry, with its dependence on plasma stopping power, can provide a practical diagnostic for NIF capsule dynamics.

\begin{acknowledgments}
We thank E.~Kemp and Z.~Walters for the design starting point, and J.~L.~Goodman and P.~A.~Bradley for insightful discussions. 
This work was supported by the U.S. Department of Energy through the Los Alamos National Laboratory. Los Alamos National Laboratory is operated by Triad National Security, LLC, for the National Nuclear Security Administration of U.S. Department of Energy (Contract No.~89233218CNA000001).
\end{acknowledgments}

\section*{Author declarations}
\noindent\textbf{Conflict of Interest}

The authors have no conflicts to disclose.

\section*{Data availability}
The data that support the findings of this study are available from the corresponding author upon reasonable request.

\bibliography{biblio}

\end{document}